# Approaches to Improving the Accuracy of Machine Learning Models in Requirements Elicitation Techniques Selection

Denys Gobov[1], Olga Solovei[2]

[1] National Technical University of Ukraine "Igor Sikorsky Kyiv Polytechnic Institute", Kyiv, Ukraine
[2] Kyiv National University of Construction and Architecture, Ukraine
`d.gobov@kpi.ua, solovey.ol@knuba.edu.ua`

**Abstract:** Selecting techniques is a crucial element of the business analysis approach planning in IT projects. Particular attention is paid to the choice of techniques for requirements elicitation. One of the promising methods for selecting techniques is using machine learning algorithms trained on the practitioners' experience considering different projects' contexts. The effectiveness of ML models is significantly affected by the balance of the training dataset, which is violated in the case of popular techniques. The paper aims to analyze the efficiency of the Synthetic Minority Over-sampling Technique usage in Machine Learning models for elicitation technique selection in case of the imbalanced training dataset and possible ways for positive feature importance selection. The computational experiment results confirmed the effectiveness of using the proposed approaches to improve the accuracy of machine learning models for selecting requirements elicitation techniques. Proposed approaches can be used to build Machine Learning models for business analysis activities planning in IT projects.

## 1. Introduction

The choice of techniques for effectively identifying requirements in developing IT solutions is essential in planning business analysis work. A thorough understanding of the variety of techniques available, their advantages and disadvantages assists the business analyst in adapting to a particular project context [1]. The business analyst must create a combination of techniques to guarantee the success of software requirements identification activities, as it is impossible to fulfill all project stakeholders' needs using just one technique [2]. One approach to solving this problem is to use machine learning models, where training samples are formed based on practitioners' experience or recommendations. For example, in studies [3,4], a machine learning model was built that recommends the usage of elicitation techniques depending on the combinations of factors. However, the use of machine learning models for commonly used techniques and the Accuracy of their work is associated with several difficulties. If we select the most frequently used techniques as a target class, the gathered dataset got imbalanced, i.e., most observations belong to a target class with a positive value equal to "1," which indicates that the technique was used in the project.

The mentioned model from [3] was constructed with Decision Jungle Tree (DJT) algorithm that was empirically selected as the most efficient. DJT learner, like a binary decision tree learner, creates a tree that is biased to the majority class when a dataset is imbalanced [5]. The reason causes are the following: while recursively partitioning the dataset so that the observations with similar target values are grouped together, it qualifies the candidate split of the node $m$ using the parameters that minimize the impurity $Q = argmin_t G(Q_m, t)$ (1), where

$$G(Q_m, t) = \frac{r_m^{left}}{r_m} H^{left}\left(Q_m^{left}(t)\right) + \frac{r_m^{right}}{r_m} H^{right}\left(Q_m^{right}(t)\right) \quad (2)$$

$H^{left}\left(Q_m^{left}(t)\right)$ (3), $H^{right}\left(Q_m^{right}(t)\right)$ (4) – are gini or entropy measures of split's impurity for the classification task. The lower the value of $H^{left}\left(Q_m^{left}(t)\right)$ (5), $H^{right}\left(Q_m^{right}(t)\right)$ (6), the better the split. When the dataset is imbalanced, there is a significant probability that the majority of the class samples are included in the same nodes, which makes entropy close to zero. As a result, the created tree is biased toward the majority class [6,7].

The mentioned problem negatively impacts the model's performance which is visible in Table "Accuracy metrics' values" in the study [3]: the high value of Accuracy (>0.9) for techniques: "Interview", "Document Analysis" and relatively low value (~0.7) of the area under the ROC curve (AUC). According to studies [8,9], AUC is an accurate indicator of the model's prediction ability when a dataset is imbalanced because it shows the ability of the classifier to rank the positive instances relative to the negative ones. The ROC curve graph is based upon true positive rate (tpr) and false positive rate (ftp), so it does not depend on the class's distributions' changes. In contrast, Accuracy measures the minimization of the overall error to which the minority class contributes very little [10].

To tackle the problem and improve the performance, in the current work, we will propose to enhance the built ML model by adding a data preprocessing step, which is to balance the dataset before training a machine learner algorithm. The prediction's quality will be measured based on the same metrics used in work [3] to obtain the predictions' comparable results.

Another point is a "Positive feature importance" algorithm used in [3] – according to which features' importance is identified based on the build-in capability of the decision tree learner when the feature used in the node's split most often receives the higher score. Such an approach depends on the learner's prediction Accuracy and can result in a wrong feature's score when a created decision tree is biased toward the majority class. In the current work, we propose to use machine learner-independent methods to identify the feature's importance score. The paper is structured as follows. Section 2 reviews the related works on approaches for handling imbalanced datasets in ML models. Section 3 contains the description of the input data and the experiment's scheme: dataset characteristics, data pre-processing procedure, train and test procedures. Section 4 is devoted to the experiments' results of using proposed improvements for the ML model. Section 5 concludes the paper with the main study findings and future work.

## 2. Related work

In the study [11], to improve classification models' results for imbalance datasets are considered the following data preprocessing strategies: 1) random resampling: 1.1) resampling the small class at random until it contains the number of samples that is equal to the majority class; 1.2) focused resampling the small class only with data occurring close to the boundaries with a factor equal to 0.25 between the concept and its negation. 2) down-sizing: 2.1) random down-sizing – eliminating at random the elements of the over-sized class until it matches the size of the majority class; 2.2) focused down-sizing - eliminating only elements further away from the boundaries; 3) learning by recognition - to use unsupervised machine learning algorithms and ignore target class in dataset. The study concluded that resampling and down-sizing methods are very effective compared to the recognition-based approach. However, it was left without a recommendation, which is a preferable method – random or focused resampling/down-sizing. Furthermore, it was specified that by random down-sizing of the majority class, there is a chance to exclude meaningful information; therefore, the preferences to be given to resampling or focused down-sizing.

In the study [12] it was proposed an over-sampling approach – Synthetic Minority Over-sampling Technique (SMOTE) in which the minority class is over-sampled by creating "synthetic" examples that are identified as k minority class nearest neighbors. If the amount of over-sampling needed is 100%, then the five nearest neighbors are chosen. The authors specified that the SMOTE approach could improve the accuracy of classifiers because, with an over-sampled dataset, the classifier builds larger decision regions. After all, the created tree isn't biased. Except for the Accuracy, oversampling algorithm SMOTE can improve the CV score, F1 Score, and recall of classifiers [13]. Further enhancements to SMOTE were proposed in the study [14] – the new method was called SMOTEBoost – according to which SMOTE is applied on each boosting round. With SMOTEBoost, the higher prediction Accuracy of the classifier was achieved.

The study [15] proposed to combine the sampling technique and ensemble idea, i.e., to make the dataset balanced by down-sizing or over-sampling and after to grow each tree. The Random Forest algorithm was used in experiments with methods: SMOTE, SMOTEBoost (called Balanced Random Forest), and the results are compared with the Weighted Random Forest algorithm. The conclusions were that both Weighted Random Forest and Balanced Random Forest outperformed other learners, and there was no winner between those two learners. The comparison was made by applying the ROC convex hull method [16]. The better the "more northwest" ROC convex hull because it corresponds to the classifier with a lower expected cost.

Several filter methods are proposed in studies [17] to have the most predictive feature selected independently from machine learning algorithms. The appropriate filter's method is selected depending on the parameters: a type of machine learning task; in case of classification – several unique values in target class (binary or multi-class); type of predictive features – continuous/discrete or categorical; type of data in the dataset – flat feature; structure feature; linked data; multi-source; multi-view; streaming feature; streaming data. However, the selection of the best-fit method for the concrete dataset was not formalized; therefore, an empirical test is required.

In the current work, based on the results of studies [11, 12, 14, 15,16], to balance a dataset, we will apply the over-sampling method SMOTE. We also use a learning algorithm called Random Forest Classifier to build a model. The mentioned enhanced methods: SMOTEBoost and Weighted Random Forest, won't be used to avoid over complication of our model. We will use the ROC convex hull method to evaluate the built model's ability to recommend an elicitation technique. However, classification task metrics like precision and recall will be calculated to compare the effect of the proposed enhancements with the results achieved in the work [3]. Because the study [17] shows no recommendations on how systematically identify the filter technique which fits best the concrete dataset, in the current study, we will empirically find an optimal for our dataset a filter method.

## 3. Methodology

To apply and evaluate the effects of the proposed enhancements, we will include additional blocks in the traditional supervised machine learning work cycle ("blue" rectangles in Fig. 1). A new block, "Data Balancing," will generate synthetic samples that will diminish the class-imbalance problem with method SMOTE. The applied technique will get a new sample $s = x + k \cdot \left( x^R - x \right)$ (7) as linear combinations of two samples from the minority class ($x^R$ and $x$) with $0 \leq k \leq 1$; $x^R$ is randomly chosen among the 5-minority class nearest neighbors of x.

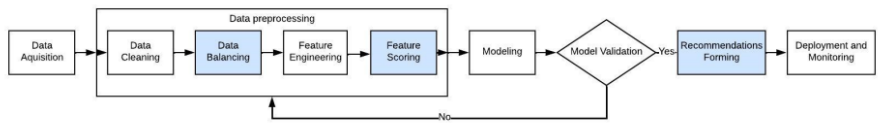

**Fig. 1.** Supervised machine learning work cycle with added blocks.

In the added new "Feature scoring" block, we use the filter methods, independent from the learning algorithm, to estimate the score of each feature. Different strategies can be used to get feature scores without knowing which filter method fits the dataset the best: 1) to average the feature's scores received by applying different filters; 2) to choose the feature's highest score from the scores received by applying different filters; 3) to apply learning algorithm with feature selected by different methods and select the best filter method to get features' scores based on the analysis of ROC convex hull. The last strategy will be used in the current work.

In a new "Recommendations Forming" block with the model's forecasts and feature scores as input information, recommendations regarding recommended techniques will be formed.

## 4. Experiment

We execute experiments to assess the effectiveness of the proposed enhancements. The effectiveness of the built machine learning model was measured via Accuracy, AUC, precision, and recall metrics. The received metrics' values are compared with the results achieved in work [3], and validations of the proposals are based on the comparison results.

Our databases' characteristics and imbalanced ratios calculated as majority-to-minority samples are specified in Table 1. The features that are included in the dataset are two types: 1). features to describe the project's context; 2) features to list all elicitation techniques used in the project [18]. The following features belong to the first type: Country, ProjectSize, Industrial Sector, Company Type, Company Size, System/Service Class, Team Distribution, Experience, Way of Work, Project Category, BA Activities (BA Only Role), and Certified. The following features belong to the second type: Benchmarking and Market Analysis, Brainstorming, Business rules analysis, Collaborative games, Data mining, Design Thinking, Interface analysis, Interviews, Observations, Process analysis, Prototyping, Reuse database and guidelines, Stakeholders list, map or Personas, Survey or Questionnaire, Document Analysis, Workshops and focus groups, Mind Mapping.

The features' information that is included in the dataset with target class names: "Elicitation", "Document Analysis", "Interface Analysis", and "Process Analysis" is the same with the following exception when the feature's name is equal to the target class, then this feature isn't included in features' list.

The data pre-processing procedure has included: 1) removing features whose values are not unique; 2) applying SMOTE to over-sample a minority class to match 100% of the majority class.

Train and test procedures included: 1) Random Forest tree classifier is trained on the sub-dataset, which is received as a result of a random split with the proportion 80/20 for train and test correspondingly. 2) Trained learner is tested with the test subset.

**Table 1.** The characteristics of datasets

| Target class name | Majority class | Minority class | Imbalance ratio | Machine learning task | Feature type | Missing values, Y/N? |
|---|---|---|---|---|---|---|
| Interviews | 282 | 41 | 6.9 | Binary classification | Discrete | N |
| Document Analysis | 276 | 47 | 5.9 | | | |
| Interface Analysis | 232 | 91 | 2.5 | | | |
| Process Analysis | 213 | 110 | 1.9 | | | |

Identification of feature's importance score: based on the dataset's parameters from Table 1 to identify the feature's score can be used methods: 1) Chi-squared stats - measures dependency between non-negative feature and class, so that irrelevant for classification feature is set a low score [19]; 2) Analysis of variance (ANOVA f-test) - measures the ratio of explained variance to unexplained variance, the higher values mean more feature's importance to a target class [20]; 3) mutual information (MI)– measures the amount of shared information between independent feature and target class. It can detect non-linear dependencies, which is its main strength, and it can be applied when the dataset includes continuous, discrete, or both types of data [21,22].

Because the types of the results of the mentioned filters are different, to understand which filter best fits datasets from Table 1, we will do the following: 1) create ROC curves calculated by Random Forest with balance data and feature selected by methods: chi-squared stats, mutual information, Anova f-value; 2) select the best-fit method based on ROC convex hull graphs analysis.

## 5.   Results and discussions

Fig. 2 illustrates how the split's impurity measured by entropy increased after the dataset was balanced by applying SMOTE, which indicates that more samples from different target classes are included in the node; hence both classes are presented before the decision in favor of the majority, or minority class is made, and the created tree isn't biased. The gap between mean entropies is ≈10% for datasets with imbalance ratios of more than 5 (Fig. 2- pictures (a) and (b)), and it is 5% and 4% for datasets with imbalance ratios of 2.5 and 1.9, correspondingly. Therefore, the graphs in Fig. 2 display the direct dependency between the decision tree's bias and the dataset's imbalance ratio.

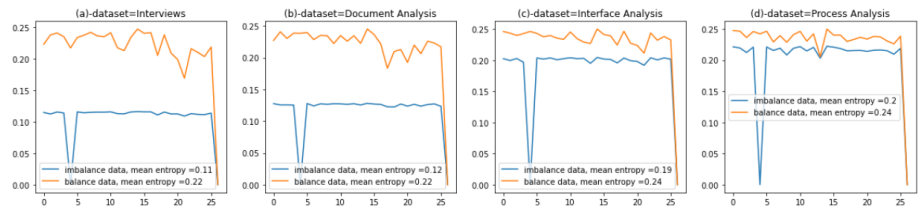

**Fig. 2.** Split's impurity of imbalance and balance datasets calculated by entropy

The values of the performance metrics of Random Forest learners with a balanced test subset are recorded in Table 2. The area under ROC curves for the balanced dataset increased significantly compared to the imbalanced dataset, and accuracy increased as well.

Table 2. Accuracy and AUC for balanced by SMOTE and original dataset

| Techniques | Accuracy | | | AUC | | |
|---|---|---|---|---|---|---|
| | Imbalanced | Balanced | Improving | Imbalanced | Balanced | Improving |
| Interviews | 0,938 | 0.94 | 0% | 0,720 | 0.97 | 35% |
| Document Analysis | 0,901 | 0.91 | 1% | 0,728 | 0.97 | 33% |
| Interface Analysis | 0,790 | 0.84 | 6% | 0,676 | 0.9 | 33% |
| Process Analysis | 0,778 | 0.82 | 5% | 0,757 | 0.88 | 16% |

Similar improvements are recorded in Fig. 3, where SMOTE-Random Forest computed ROC convex hull colored by orange, and Random Forest computed ROC convex hull shown by blue color with the imbalanced dataset. The SMOTE-Random Forest ROC convex hull dominates the ROC convex hull without SMOTE, making SMOTE a more optimal classifier.

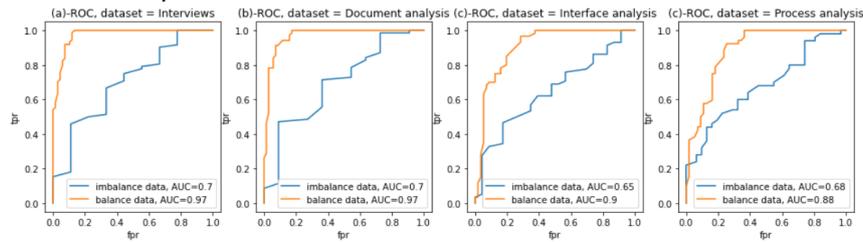

Fig. 3. ROC convex hull of imbalance and balance datasets calculated by Random Forest classifier

The p-value of paired t-test for precision and recall received for the balanced dataset and imbalanced dataset from [3] are recorded in Table 3. p-value= 0.018 indicates a statistically significant difference in the precision of the Random Forest classifier with a balanced dataset compared to imbalance which means that the ability of the classifier not to label as positive a sample that is negative is affected when the negative's class distribution's changes because of over-sampling. The p-value=0.087 indicates the statistically insignificant difference in the recall, i.e., the classifier's ability to correctly find all positive samples when the positive class is a majority class has not been affected by applying the oversampling of the minority class.

Table 3. p-value of paired t-test of precision and recall

| Dataset | Interviews | Document Analysis | Interface Analysis | Process Analysis | p-value |
|---|---|---|---|---|---|
| Precision, balanced data | 0.89 | 0.88 | 0.81 | 0.78 | 0.018 |
| Precision, imbalanced data | 0.936 | 0.899 | 0.831 | 0.818 | |
| Recall, balanced data | 1 | 0.96 | 0.9 | 0.88 | |
| Recall, imbalanced data | 1 | 1 | 0.922 | 0.9 | 0.087 |

To select the best feature selection method, we used ROC convex hull graphs presented in Fig. 4. For each dataset from Table 1, graphs show the "more northwest" ROC curve when features are selected by the mutual information method. Features' importance score for each dataset calculated by the mutual information method is recorded in Tables 4-5. To type of recommendations can be produced based on the results in Tables 4-5: 1). Collaborating filtering - the list of elicitations techniques that can be used in the project in addition to the elicitation technique that is predicted by a machine learning model. 2). Content-based filtering – based on the similarity of the project's context the system can recommend the elicitation technique to be used. Due to form the recommendations of the first and the second types, the system selects from Tables 4-5 the features whose score is higher than a particular value and which correspondingly belong to the first and the second feature's type. The value of the score, which the system uses to select the feature, is proposed to be a user's choice.

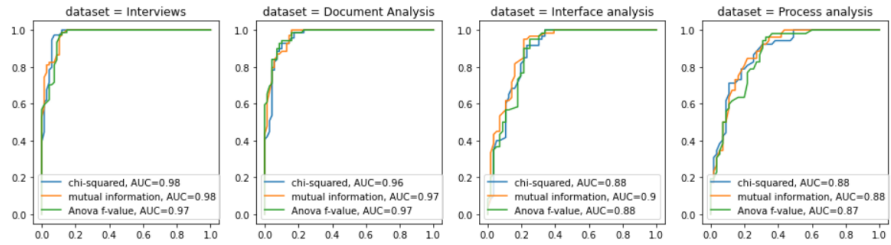

Figure 4. ROC convex hull calculated by Random Forest with balance data and feature selected by methods: chi-squared, mutual information, Anova f-value

Table 4. Features' importance score. Interviews and Document analysis.

| Interviews | | Document analysis | |
|---|---|---|---|
| Feature | Score | Feature | Score |
| Project Size | 0.3 | Experience | 0.32 |
| Experience | 0.28 | Project Size | 0.27 |
| WoW | 0.27 | WoW | 0.25 |
| Prototyping | 0.25 | Industrial Sector | 0.22 |
| Project Category | 0.23 | Company Size | 0.2 |
| Company Type | 0.21 | Project Category | 0.2 |
| Process analysis | 0.19 | Process analysis | 0.19 |
| Industrial Sector | 0.18 | Company Type | 0.19 |
| Company Size | 0.18 | Interface analysis | 0.18 |
| Interface analysis | 0.17 | Observations | 0.17 |
| Brainstorming | 0.16 | BA Only Role | 0.16 |
| Observations | 0.16 | Business rules analysis | 0.15 |
| Workshops and focus groups | 0.15 | Workshops and focus groups | 0.14 |
| Business rules analysis | 0.15 | Reuse database and guidelines | 0.14 |
| Reuse database and guidelines | 0.14 | Benchmarking and Market Analysis | 0.14 |

| Feature | Score | Feature | Score |
|---|---|---|---|
| Document analysis | 0.14 | Prototyping | 0.13 |
| Team Distribution | 0.13 | Stakeholders list, map or Personas | 0.13 |
| Stakeholders list, map or Personas | 0.13 | Interviews | 0.12 |
| System/Service Class | 0.13 | Survey or Questionnaire | 0.12 |
| BA Only Role | 0.13 | Brainstorming | 0.12 |
| Survey or Questionnaire | 0.12 | Design Thinking | 0.12 |
| Data mining | 0.11 | System/Service Class | 0.11 |
| Benchmarking and Market Analysis | 0.11 | Data mining | 0.1 |
| Certified | 0.09 | Team Distribution | 0.1 |
| Design Thinking | 0.08 | Certified | 0.08 |
| Collaborative games | 0.06 | Collaborative games | 0.03 |
| Mind Mapping | 0 | Mind Mapping | 0.01 |

**Table 5.** Features' importance score. Interface analysis and Process Analysis

| Interface analysis | | Process Analysis | |
|---|---|---|---|
| Feature | Score | Feature | Score |
| Experience | 0.22 | Business rules analysis | 0.15 |
| Project Size | 0.15 | Project Category | 0.15 |
| WoW | 0.15 | Company Type | 0.14 |
| Company Type | 0.15 | Industrial Sector | 0.13 |
| Project Category | 0.14 | WoW | 0.12 |
| Team Distribution | 0.14 | Experience | 0.11 |
| Observations | 0.13 | Workshops and focus groups | 0.11 |
| Prototyping | 0.12 | Stakeholders list, map or Personas | 0.1 |
| Document analysis | 0.12 | Benchmarking and Market Analysis | 0.1 |
| Company Size | 0.12 | Team Distribution | 0.09 |
| Brainstorming | 0.11 | BA Only Role | 0.09 |
| BA Only Role | 0.11 | Observations | 0.08 |
| Process analysis | 0.11 | Data mining | 0.07 |
| Business rules analysis | 0.1 | Reuse database and guidelines | 0.07 |
| Workshops and focus groups | 0.1 | Survey or Questionnaire | 0.07 |
| Stakeholders list, map or Personas | 0.1 | Project Size | 0.07 |
| Benchmarking and Market Analysis | 0.09 | Prototyping | 0.06 |
| Design Thinking | 0.07 | Certified | 0.05 |
| Survey or Questionnaire | 0.07 | Interface analysis | 0.05 |
| Data mining | 0.06 | Company Size | 0.04 |
| Industrial Sector | 0.05 | Design Thinking | 0.03 |
| Reuse database and guidelines | 0.05 | Document analysis | 0.03 |
| Interviews | 0.04 | Interviews | 0.03 |
| Certified | 0.04 | System/Service Class | 0.02 |
| System/Service Class | 0.03 | Mind Mapping | 0.01 |
| Mind Mapping | 0.02 | Brainstorming | 0.01 |
| Collaborative games | 0 | Collaborative games | 0 |

## 6. Conclusions

The current study proposes several enhancements to the machine learning model for getting recommendations on elicitation techniques usage based on the combinations of factors. The first proposal is to balance the dataset by applying SMOTE methods before the training and testing classifier. As a result, improvement in the "Accuracy" indicator was obtained in the range from 0% to 6%, and the "AUC" indicator from 16% to 35%. The dependency between the value of the imbalance ratio and created tree biased to the majority class was visual. The results can be used in future works to formalize whether applying SMOTE to balance a dataset is necessary, as an increased dataset will demand more operation time for machine learning. The second proposal is to calculate the features' importance score independently from the machine learner classifier. A mutual information method was identified as the best fit for the considered datasets. The produced list of the most predictable features is proposed to use in recommended systems of both types: collaborating filtering and content-based filtering. The results obtained allow the construction of more precise models for the recommendation of requirements elicitation techniques contingent upon the project context. Further research can be conducted to explore additional BA tasks to determine correlations and make suggestions for selecting techniques for requirement specification and modeling, validation, and verification based on balanced datasets.